\documentstyle[prb,aps,multicol]{revtex}
\input{psfig}

\begin{document}
\title{Signatures of polaronic excitations in quasi-one-dimensional LaTiO$_{3.41}$}
\author{C. A. Kuntscher$^{*}$}
\address{Material Science Center, University of Groningen, Nijenborgh 4,
9747 AG Groningen, The Netherlands \\
and 1. Physikalisches Institut, Universit\"at Stuttgart, Pfaffenwaldring 57, 
D-70550 Stuttgart, Germany}

\author{D. van der Marel}
\address{Material Science Center, University of Groningen, Nijenborgh 4,
9747 AG Groningen, The Netherlands}

\author{M. Dressel}
\address{1. Physikalisches Institut, Universit\"at Stuttgart, Pfaffenwaldring 57, 
D-70550 Stuttgart, Germany}

\author{F. Lichtenberg and J. Mannhart}
\address{Experimentalphysik VI, Institut f\"ur Physik, EKM, 
Universit\"at Augsburg, Universit\"atsstr. 1, D-86135 Augsburg, Germany}

\date{\today}                                      
\maketitle

\begin{abstract}
The optical properties of quasi-one-dimensional metallic LaTiO$_{3.41}$ 
are studied for the polarization along the $a$ and $b$ axes. With decreasing 
temperature modes appear along both directions suggestive for a phase 
transition. The broadness of these modes along the conducting 
axis might be due to the coupling of the phonons to low-energy electronic 
excitations across an energy gap. We observe a pronounced midinfrared band 
with a temperature dependence consistent with (interacting) polaron models. 
The polaronic picture is corroborated by the presence of strong electron-phonon 
coupling and the temperature dependence of the dc conductivity. 
\end{abstract}


\begin{multicols}{2}
\columnseprule 0pt \narrowtext 
Titanium oxide compounds have been investigated extensively over the last 
decades, at the latest since the discovery of high-$T_c$ superconductivity 
in the cuprates, to study the role of electronic correlations and to 
explore the doping induced transition from a Mott insulator to a metal, 
like, e.g., in La$_{1-y}$Sr$_y$TiO$_3$. \cite{Fujimori92,Hays99} Of 
further interest are the titanates because of the proposed polaronic 
nature of their charge carriers. For example, the existence of small 
polarons in La$_{1-\gamma}$TiO$_{3\pm\delta}$ was shown by dc resistivity 
and thermoelectric power measurements.\cite{Hays99,Jung00}
Signatures of polaronic carriers were also found in the optical 
response of TiO$_2$, BaTiO$_3$, and SrTiO$_3$ in the form of 
a midinfrared (MIR) 
band.\cite{Bogomolov68,Gerthsen65,Eagles84,Calvani93,Kabanov95} 
Since a MIR band of (spin-) polaronic origin at $\approx$1000 cm$^{-1}$ was 
also found in several cuprates \cite{Thomas92,Calvani96,Lupi99,Gruninger99} 
it seems to be a characteristic feature of this class of compounds for
low doping as well. The proposal that (bi)polarons might play a major role 
for the high-$T_c$ superconductivity\cite{Emin94} stimulated a vast amount 
of experimental investigations on this issue.

On the other hand, the nature of the polaronic carriers in the titanates
is still under discussion. In this paper we present the optical properties
of another titanium oxide compound, LaTiO$_{3.41}$, to search for polaronic 
signatures and test their compatibility with the existing models.
Indeed, we find a strong MIR band in the optical response showing a strong
temperature dependence. A particular property of LaTiO$_{3.41}$
is its quasi-one-dimensional (quasi-1D) metallic character which was 
recently found by resistivity measurements. \cite{Lichtenberg01}
It is interesting to note that a MIR band was also observed for a variety 
of organic and anorganic quasi-1D metals, like TTF-TCNQ,\cite{Hinkelmann75}
the Bechgaard salts,\cite{Jacobsen83} $\beta$-Na$_{0.33}$V$_2$O$_5$,\cite{Presura02}
and SrNbO$_{3.41}$.\cite{Kuntscher02} Whether this band is related to the
1D transport and, in particular, is of (vibrational or spin-) polaronic 
origin, are open questions which we want to address here.
\begin{figure}[h]
\centerline{\psfig{file=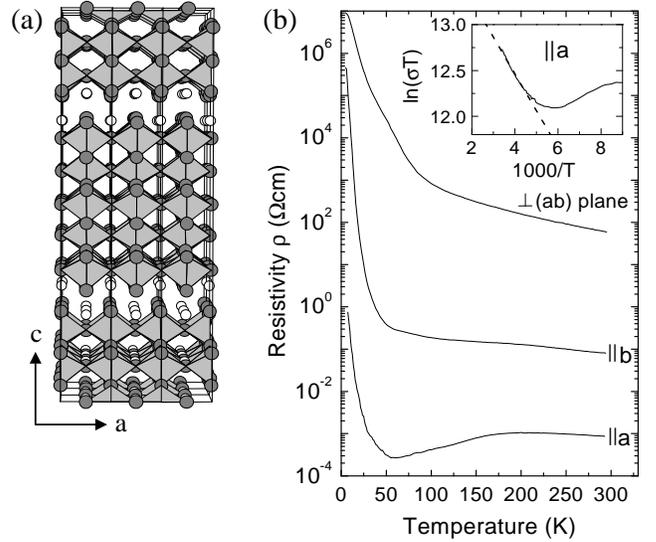,width=85mm,clip=}}
\caption{(a) Projection along the $b$ axis of the schematic LaTiO$_{3.4}$  
crystal structure. The TiO$_6$ octahedra (light grey) are connected 
continuously via their apical sites forming 1D chains along $a$ 
(grey circles: O atoms, white circles: La atoms; Ti atoms hidden 
within the TiO$_6$ octahedra). 
(b) dc resistivity $\rho$ of LaTiO$_{3.41}$ versus temperature $T$ 
along the three crystal axes. 
Inset: Fit (dashed line) of $\sigma T$ between 200 and 300 K according 
to Eq.\ (\ref{hopping}) for small polaron hopping.}
   \label{fig:resist}
\end{figure}
\begin{figure}[t]
\centerline{\psfig{file=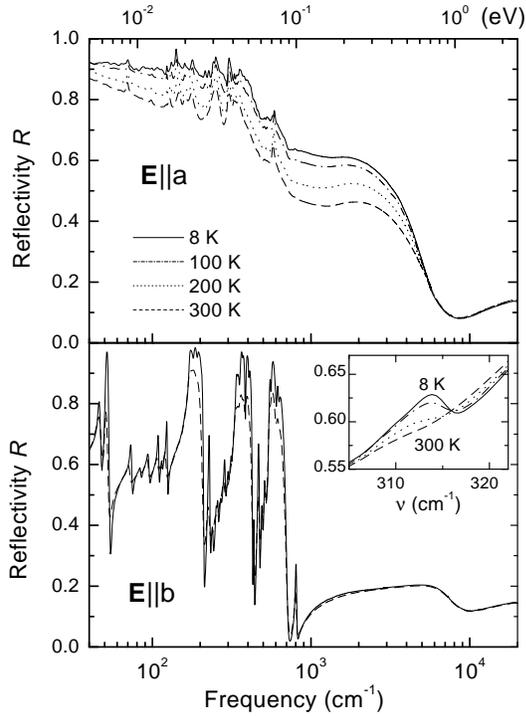,width=70mm,clip=}}
\caption{Reflectivity $R$ of LaTiO$_{3.41}$ for 
\textbf{E}$\parallel$$a$ and \textbf{E}$\parallel$$b$.
Inset: Detailed temperature dependence of a phonon mode for 
\textbf{E}$\parallel$$b$ appearing below 200 K.}
\label{fig:reflec}
\end{figure}

Single crystals of LaTiO$_{3.41}$ were grown by a floating 
zone melting process;\cite{Lichtenberg01} the
oxygen content was determined by thermogravimetry. The layered crystal
structure [see Fig.\ \ref{fig:resist}(a)] is built from slabs of distorted 
TiO$_6$ octahedra parallel to the ($a$,$b$) plane; it is monoclinic with 
lattice constants $a$=7.86 \AA, $b$=5.53 \AA, $c$=31.48 \AA, and 
$\beta$=97.1$^{\circ}$. Like for the closely related compound SrNbO$_{3.41}$ 
a characteristic property of the structure are the 1D chains of octahedra along 
the $a$ axis.\cite{Lichtenberg01}
LaTiO$_{3.41}$ belongs to the homologous series La$_n$Ti$_n$O$_{3n+2}$
which exhibits a very rich phase diagram:\cite{Lichtenberg01}
stoichiometric LaTiO$_{3.00}$ ($n$=$\infty$, Ti $3d^1$) is an 
antiferromagnetic Mott insulator, while for off-stoichiometric 
La$_{1-\gamma}$TiO$_{3\pm\delta}$ metallic character of polaronic 
type was observed;\cite{Hays99,Jung00} LaTiO$_{x}$ for 
3.00$<$$x$$<$3.50 is conducting,\cite{Lichtenberg01}
and in particular LaTiO$_{3.41}$ (Ti $3d^{0.18}$) shows a quasi-1D 
metallic behavior;\cite{Lichtenberg01} LaTiO$_{3.50}$ ($n$=4, 
Ti $3d^0$) is a ferroelectric band insulator.\cite{Nanamatsu74}
The dc resistivity as a function of temperature 
$T$ depicted in Fig.\ \ref{fig:resist}(b) demonstrates the strongly 
anisotropic character of LaTiO$_{3.41}$. 
Along the $b$ direction and perpendicular to the ($a$,$b$) plane the 
compound shows a semiconducting behavior, whereas along the $a$ direction 
the $T$ dependence is more complicated: Below 60 K $\rho_a$ rises steeply 
with decreasing $T$, between 60 and 200 K its $T$ dependence is metallike, 
and above 200 K $\rho_a$ slightly decreases with increasing $T$. This 
temperature dependence will be discussed in more detail later.

Near-normal incidence reflectivity spectra were measured from 40 to 
6000 cm$^{-1}$ (5 meV-0.74 eV) utilizing a Fourier-transform
spectrometer equipped with an ultra- 
\begin{figure}[t]
\centerline{\psfig{file=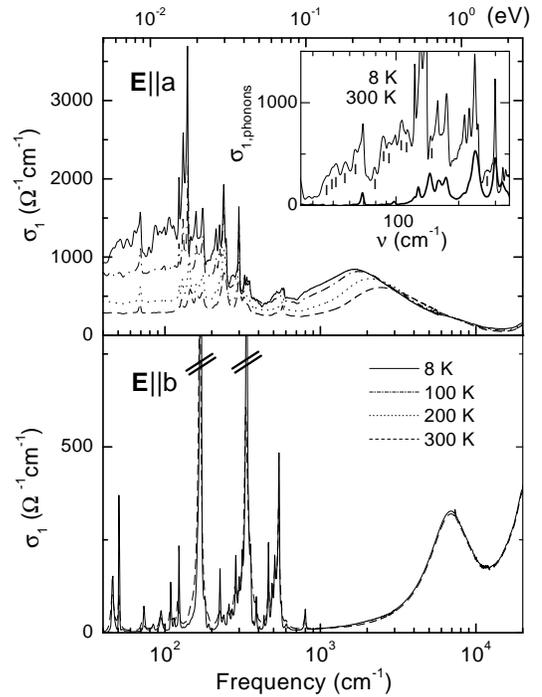,width=70mm,clip=}}
\caption{Optical conductivity $\sigma_1$ of 
LaTiO$_{3.41}$ for \textbf{E}$\parallel$$a$ and \textbf{E}$\parallel$$b$.
Inset: Phonon contribution $\sigma_{1,\rm phonons}$ to $\sigma_1$
for \textbf{E}$\parallel$$a$ from 35 to 350 cm$^{-1}$,
obtained by subtracting the Drude term and the MIR band
(from the Drude-Lorentz fit) from the measured spectrum; ticks 
indicate the broad modes which appear below 100 K.}
\label{fig:cond}
\end{figure}
stable optical cryostat. To obtain
absolute reflectivities the spectra were divided by the reflectivity
spectra of a gold film recorded at the same set of temperatures,
where the film was deposited {\it in situ} on the sample. Since
no temperature dependence was found above 6000 cm$^{-1}$, the spectra
were extended to 36 000 cm$^{-1}$ by the room-temperature (RT) 
reflectivity data recorded with a variable angle spectroscopic ellipsometer. 
Each reflectivity spectrum was extrapolated to low and high frequencies 
according to a Drude-Lorentz fit; from the subsequent Kramers-Kronig 
transformation the phase of the reflection coefficient and thus the 
complex dielectric function could be obtained.

\begin{figure}[h]
\centerline{\psfig{file=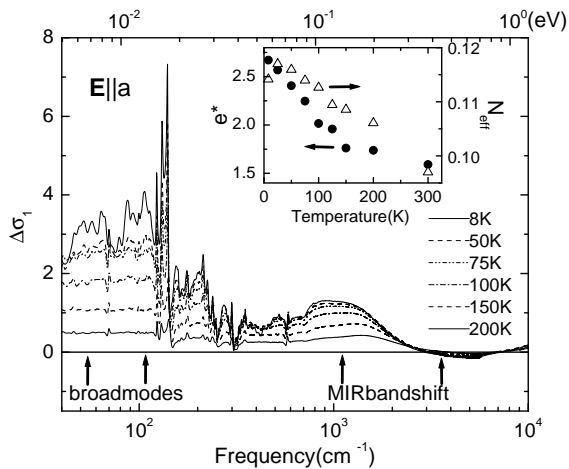,width=75mm,clip=}}
\caption{Normalized difference $\Delta\sigma_1$ [Eq.\ (\ref{sigmadiff})] 
of LaTiO$_{3.41}$ for \textbf{E}$\parallel$$a$, illustrating the appearance 
of broad modes and the MIR band shift with decreasing temperature. 
Inset: Effective charge $e^*$ of the phonon modes and effective carrier 
number $N_{\rm eff}$ per Ti atom of the MIR band as a function of 
temperature.}
\label{fig:delta}
\end{figure}

The polarization-dependent reflectivity spectra $R(\omega)$ of LaTiO$_{3.41}$ 
for several temperatures are shown in Fig.\ \ref{fig:reflec}.
They clearly demonstrate its quasi-1D metallic character:
For the polarization \textbf{E}$\parallel$$a$, i.e., along the chains, 
we find a metallic behavior with a high reflectivity at low frequencies 
and a sharp plasma edge ($\omega_p$$\approx$3600 cm$^{-1}$), while the 
overall low reflectivity for 
\textbf{E}$\parallel$$b$ indicates an insulating character. 
The optical conductivity $\sigma_1(\omega)$ is presented in Fig.\ \ref{fig:cond}. 
For \textbf{E}$\parallel$$b$ no electronic absorption is observed in the 
far infrared (FIR); the spectrum below 1000 cm$^{-1}$ consists 
of a large number of phonon lines, and the onset of interband transitions 
is found above 1000 cm$^{-1}$. Along the chains $\sigma_1$ consists 
of a Drude peak superimposed by vibrational lines and a pronounced MIR 
band.\cite{comment0}
With decreasing temperature we observe two major changes in the 
\textbf{E}$\parallel$$a$ conductivity: (i) in the FIR region, between 
50 and 300 cm$^{-1}$, modes (indicated by ticks in the inset of 
Fig.\ \ref{fig:cond}) possibly of vibrational origin appear below 
100 K which are broader than the others, and (ii) the MIR 
band, located around 2500 cm$^{-1}$ at RT, continuously shifts to lower 
frequencies and its intensity increases; besides, an additional feature 
appears around 940 cm$^{-1}$. The changes are illustrated by the 
normalized difference,
\begin{equation}  \label{sigmadiff}
\Delta\sigma_1(\omega,T)=[\sigma_1(\omega,T)-\sigma_1(\omega,300 $K$)]/\sigma_1(\omega,300 $K$),
\end{equation}
shown in Fig.\ \ref{fig:delta}. In the FIR, below 150 cm$^{-1}$, 
for $T$$\geq$100 K $\Delta\sigma_1(\omega)$ has slope zero and it 
increases with decreasing $T$ due to the increase of $\sigma_{dc}$; 
but below 100 K the slope clearly changes and in particular for 
45--100 cm$^{-1}$ $\Delta\sigma_1$ increases strongly, indicating 
additional excitations in this frequency range.

\begin{figure}[t]
\centerline{\psfig{file=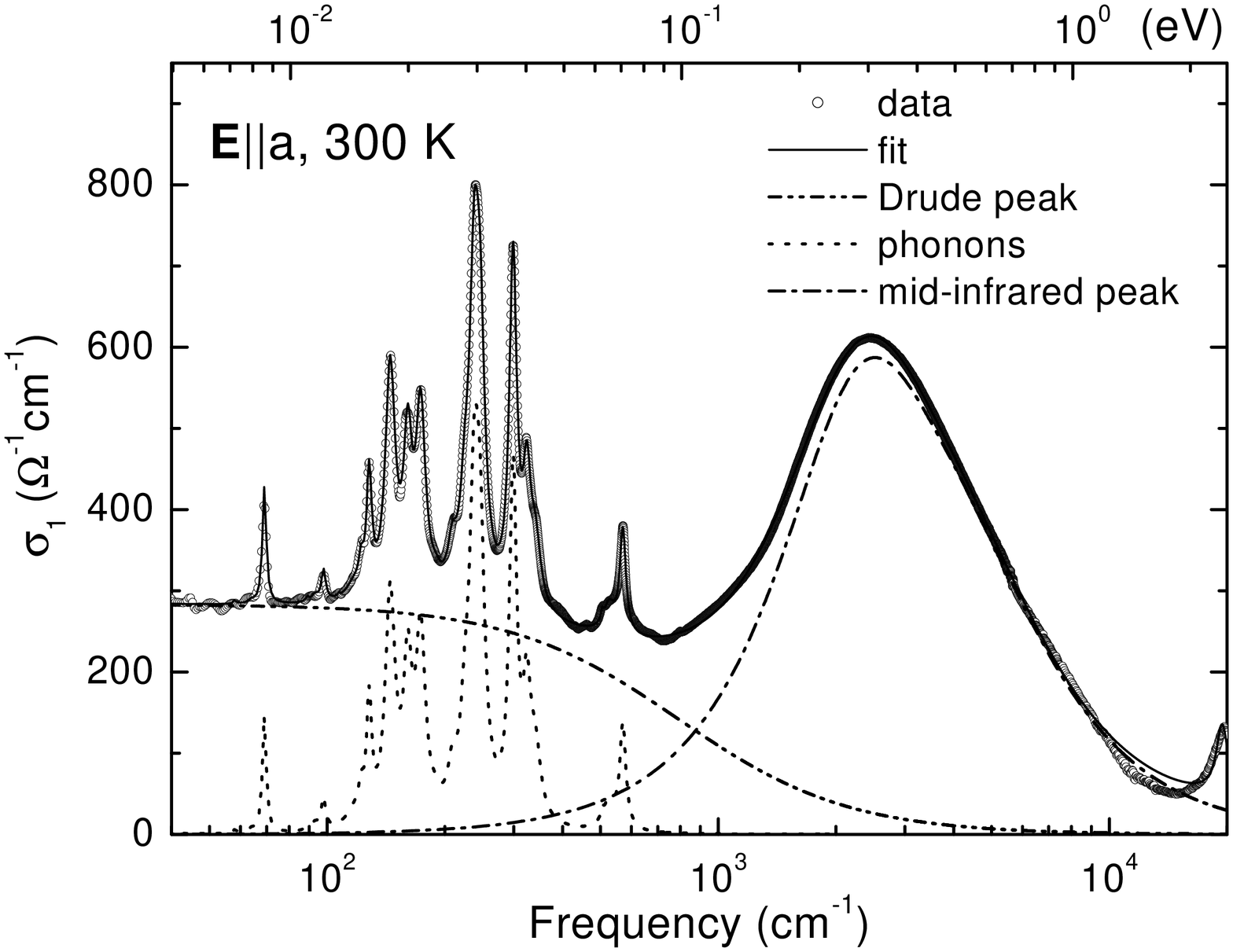,width=75mm,clip=}}
\caption{Fit of the RT conductivity $\sigma_1$ of LaTiO$_{3.41}$
with the Drude-Lorentz model.}
    \label{fig:fit}
\end{figure}

To quantify these changes we fitted the \textbf{E}$\parallel$$a$ 
conductivity spectra with the Drude-Lorentz model. The MIR band was 
modelled by two Lorentzian functions. As an example we show the fit 
for the RT spectrum in Fig.\ \ref{fig:fit}. 
This way we were able to extract the various contributions 
(Drude peak, ``phonon'' modes, MIR band) for each temperature. 
From the spectral weights we calculated the effective number
of carriers (per Ti atom) for each contribution according to
$N_{\rm eff}=[2m_eV/(\pi e^2)]/20 \int_0^{\omega_m}\sigma_1(\omega)d\omega$,
where $m_e$ is the free electron mass, and $V$ the unit cell volume
(containing 20 formula units). The upper integration limit $\omega_m$ 
was set to 15000 cm$^{-1}$, where the onset of the interband transitions
is located. 
As is shown in the inset of Fig.\ \ref{fig:delta}, the effective carrier 
number of the MIR absorption increases as the temperature decreases, but 
below 75 K a saturation sets in. 
From the integrated intensities of the phonon modes and by using the ionic 
charges of the atoms we calculated the effective charge $e^*$ according
to the sum rule.\cite{Dressel02} The temperature dependence of $e^*$ 
(inset of Fig.\ \ref{fig:delta}) illustrates the substantial spectral 
weight growth of the modes with decreasing temperature. 

We now want to discuss the observed changes in the optical conductivity 
in more detail.
(i) The appearance of vibrational modes in general suggests 
a breaking of the crystal symmetry. Some of the new modes
for \textbf{E}$\parallel$$a$ are already very weakly present  
in the 125-K spectrum, but the substantial growth of their spectral 
weight sets in below 100 K. The occurrence of a phase transition in
LaTiO$_{3.41}$ is corroborated by the development of ten phonon lines 
for \textbf{E}$\parallel$$b$ with decreasing temperature (an example 
is shown in the inset of Fig.\ \ref{fig:reflec}).
A remarkable finding is the broadness of the new lines along the
conducting $a$ axis (see inset of Fig.\ \ref{fig:cond}) which might 
be due to the coupling of the phonons to electronic excitations. 
These low-energy excitations might be excitations across an energy
gap at $E_F$ which develops below 100 K due to the phase transition. 
A coupling of the phonons to these excitations could also explain 
their significant spectral weight growth below this temperature. 
Since the broad modes appear at frequencies above 50 cm$^{-1}$, one 
could estimate the size of the gap to $\approx$6 meV.
The appearance of strong modes along the metallic axis caused by a 
coupling of phonon bands to the electronic density was in fact 
predicted by Rice \cite{Rice76} when considering organic linear-chain 
conductors which undergo a charge-density wave transition.
For LaTiO$_{3.41}$ the presence of strong electron-phonon 
coupling is indicated by the enhanced effective charge $e^*$=1.6 at 
RT (see inset of Fig.\ \ref{fig:delta}) compared to the value 
$e^*$=1 if there is no coupling. In general, the interaction between 
a discrete level and a continuum of states gives rise to asymmetric 
peaks in the excitation spectra.\cite{Fano61} Unfortunately, the large 
electronic background (Drude contribution) and the strong overlap 
of the modes render a line-shape analysis and thus a proof of the 
asymmetry of the new modes in LaTiO$_{3.41}$ difficult.

Within this scenario two open questions remain, however: 
First, the dc resisitivity of LaTiO$_{3.41}$ shows no anomaly and 
thus no clear signature of a phase transition over the whole 
measured temperature range; a possible explanation could be the small 
number of carriers involved. Second, there is no {\it direct} evidence 
in the optical conductivity spectrum for excitations across an energy 
gap. 
Based on the present experimental results the occurrence of a phase 
transition therefore remains speculative. Additional experimental
investigations, for instance x-ray diffraction or neutron-scattering
measurements at low temperature, would be needed to clarify this issue.

(ii) The strong temperature dependence of the MIR band, namely its 
shift to lower frequencies and intensity increase for decreasing $T$ 
with a saturation at a specific temperature, renders an interpretation
of the band in terms of interband transitions unlikely. The observed
evolution with temperature is similar to that of the MIR absorption 
in the cuprate superconductors which was attributed to polaronic
excitations.\cite{Calvani96,Lupi99} The formation of polarons in
LaTiO$_{3.41}$ is corroborated by the presence of strong electron-phonon
coupling indicated by the enhanced effective charge $e^*$. 
A temperature-dependent MIR band was also found for the nickelates 
La$_{2-x}$Sr$_x$NiO$_{4+\delta}$, where it was explained by 
photon-assisted hopping of small polarons:\cite{Calvani96b,Bi93a} 
photons in the MIR range excite self-trapped carriers from a localized 
state to localized states at neighboring sites, and the absorption is 
peaked at an energy twice the polaron binding energy $E_P$.\cite{Emin93}
The position of the MIR band thus provides an estimate of $E_P$, which
yields $E_P$$\approx$$\hbar$$\omega/2$$\approx$155 meV for LaTiO$_{3.41}$
at RT.
A conduction mechanism due to the hopping of small polarons, as suggested
by the existence of a temperature-dependent MIR band, is corroborated
by the $T$ dependence of $\rho_a$ (see Fig.\ \ref{fig:resist}):
Starting from the lowest temperature, the steep drop of $\rho_a$
denotes the thermal activation of charge carriers. At $\approx$60 K
the carriers are free and with further temperature increase $\rho_a$ 
rises since the carriers are increasingly scattered by phonons; 
metallic transport is observed. The decrease of $\rho_a$ above 200 K 
indicates that the contribution due to hopping of polaronic carriers 
prevails. For the dc conductivity $\sigma(T)$ due to small polaron 
hopping one expects the thermally activated form \cite{Mott71}
\begin{equation}  \label{hopping}
T\sigma(T)\propto \exp[-E_H/(k_BT)],
\end{equation}
where $E_H$ is the hopping energy; the disorder energy is omitted since 
it is negligibly small compared to $E_H$ in crystalline bulk materials. 
In the range 200-300 K $\sigma_a(T)$ can be fitted according
to Eq.\ (\ref{hopping}) (see inset of Fig.\ \ref{fig:resist}) which 
yields $E_H$$\approx$35 meV. This thermal activation energy is a factor 
of $\approx$2.2 smaller than the activation energy 
$E_A$=$E_P/2$$\approx$78 meV,\cite{Mott71} obtained from our optical 
data; such a discrepancy between the thermal and optical activation 
energies is expected since in the thermally activated process the 
lattice has time to relax and $E_P$ is thus reduced.

On the other hand, the temperature dependence of the MIR band in 
LaTiO$_{3.41}$ -- the {\it softening} with decreasing temperature -- 
is not compatible with a noncorrelated small polaron model for strong 
electron-phonon coupling.\cite{Bogomolov68,Calvani96b,Crandles93} 
In the case of the high-$T_c$ cuprate superconductors the MIR band 
softening was interpreted in terms of large polaron models including
polaron-polaron interactions.\cite{Lupi99,Fratini98,Cataudella99,Tempere01} 
A different approach was recently used by Fratini {\it et al.}\cite{Fratini01}
who considered small polaron absorption for intermediate 
electron-phonon coupling: Their calculated optical conductivity
exhibits a noticeable transfer of spectral weight from high to low 
frequencies with decreasing temperature, like we observe in 
LaTiO$_{3.41}$. Besides, the calculated spectra show anomalous peaks
at frequencies comparable to the phonon frequencies with a strongly 
temperature-dependent intensity. These peaks, denoted as 
``polaron interband transitions'', are purely electronic in nature 
and due to transitions between different subbands in the polaron 
excitation spectrum. They might serve as an alternative explanation 
for the broad peaks found in LaTiO$_{3.41}$ along the conducting 
direction. 

Whether one of these polaron models actually applies to LaTiO$_{3.41}$, 
may be judged from a detailed comparison of the observed features 
(MIR absorption features, broad modes) to the theoretical 
absorption spectra, thereby considering that not only vibrational but 
also spin degrees of freedom could be involved in the polaron formation, 
as it was proposed for other materials with strong electronic correlations 
like the cuprates \cite{Gruninger99} or the manganites.\cite{Millis96} 
The clarification of these issues is of general importance, since a band 
in the MIR frequency range is a characteristic feature in the optical 
spectrum of quasi-1D conductors.\cite{Tajima00} According to the 
present results for LaTiO$_{3.41}$ the MIR band contains important 
information on the transport mechanism in quasi-1D systems and the 
careful study of its temperature and doping dependence is therefore 
worthwhile. 

In summary, we studied the polarization dependent optical response of 
quasi-1D metallic LaTiO$_{3.41}$ as a function of temperature. With 
decreasing temperature modes appear along both studied crystal axes 
suggestive for a phase transition. The new modes found along the 
conducting direction $a$ below 100 K are broader than the other modes, 
which might be caused by the coupling of the phonons to low-energy 
electronic excitations across an energy gap.
The \textbf{E}$\parallel$$a$ conductivity spectrum contains a pronounced 
MIR band whose temperature dependence is similar to that of the MIR 
absorption in the cuprate superconductors and consistent with 
(interacting) polaron models. The polaron formation in LaTiO$_{3.41}$ 
is corroborated by the presence of strong electron-phonon coupling and 
the temperature dependence of the dc conductivity. The findings for 
LaTiO$_{3.41}$ suggest the general importance of polaronic quasiparticles 
for the transport mechanism in quasi-1D conductors.

\subsection*{ACKNOWLEDGEMENTS}
We thank P. Haas for providing additional data.
This work was supported by the BMBF (project No. 13N6918/1), and the 
Deutsche Forschungsgemeinschaft.

\end{multicols}

\begin{references}
\bibitem{comment0}
The discrepancy between the dc conductivity $\sigma_{dc}$ and the 
low-frequency limit of the optical conductivity $\sigma_1$ along the 
chain direction $a$, most obvious below 50 K, are probably due 
to the low-frequency extrapolations used for the Kramers-Kronig 
analysis, which are based on the Drude-Lorentz fits of the 
reflectivity spectra in the range 40-36 000 cm$^{-1}$.
\end{references}
\end{document}